\documentclass[prd,showpacs,nofootinbib]{revtex4}
\usepackage{amsmath}
\usepackage{amsfonts}
\usepackage{amssymb}
\usepackage{bm}

\begin{document}

\title{An Investigation of the $K_{F}$-type Lorentz-Symmetry Breaking Gauge
Models in $N=1$-Supersymmetric Scenario. }
\author{H. Belich $^{a,d,e}$, G. S. Dias$^{b,f}$, J.A. Helay\"{e}l-Neto$%
^{c,d}$, F.J.L. Leal$^{b,c},$W. Spalenza$^{g}$ }
\affiliation{$^{a}${\small {Universidade Federal do Esp\'{\i}rito Santo (UFES),
Departamento de F\'{\i}sica e Qu\'{\i}mica, Av. Fernando Ferrari S/N, Vit%
\'{o}ria, ES, CEP 29060-900, Brasil,}}}
\affiliation{$^{b}${\small {Instituto Federal do Esp\'{\i}rito Santo (IFES),
Coordenadoria de F\'{\i}sica, Av. Vit\'{o}ria 1729, Jucutuquara, Vit\'{o}ria
- ES, 29040-780, CEP 29040-780, Brasil, }}}
\affiliation{{\small {~}}$^{c}${\small {CBPF - Centro Brasileiro de Pesquisas F\'{\i}%
sicas, Rua Xavier Sigaud 150, Rio de Janeiro, RJ, CEP 22290-180, Brasil,}}}
\affiliation{$^{d}${\small {Grupo de F\'{\i}sica Te\'{o}rica Jos\'{e} Leite Lopes, C.P.
91933, CEP 25685-970, Petr\'{o}polis, RJ, Brasil,}}}
\affiliation{$^{e}${\small {International Institute of Physics, Universidade Federal de
Rio Grande do Norte, av. Odilon Gomes de Lima 1722, CEP 59078-400, Natal-RN,
Brasil,}}}
\affiliation{$^{f}${\small {Department of Physics, University of Alberta, Edmonton,
Alberta, Canada T6G 2J1,}}}
\affiliation{$^{g}${\small Instituto Federal do Esp\'{\i}rito Santo (Ifes) - Campus
Cariacica Rod. Gov. Jos\'{e} Sette s/n , Cariacica, ES, Brasil, cep 29150-410, Vit\'{o}ria, ES, Brasil.}}
\email{belichjr@gmail.com, gilmar@ifes.edu.br, helayel@cbpf.br, nandojll@ig.com.br, spalenza@ifes.edu.br }

\begin{abstract}
In this work, we present two possible venues to accomodate the $K_{F}$-type
Lorentz-symmetry violating Electrodynamics in an $N=1$-supersymmetric
framework. A chiral and a vector superfield are chosen to describe the
background that signals Lorentz-symmetry breaking. In each case, the $\
K_{\mu \nu \kappa \lambda }$-tensor is expressed in terms of the components
of the background superfield that we choose to describe the breaking. We
also present in detail the actions with all fermionic partners of the
background that determine $\ K_{\mu \nu \kappa \lambda }$.
\end{abstract}

\maketitle

\section{Introduction}

The Standard Model Extension (SME) is the natural framework to investigate
properties of Lorentz-violation in physical systems involving possible
extensions of Higgs mechanism. The SME is by-now a very estimulating
research area and a great deal of results have been attained which include
bounds on the photon mass \cite{photons1}, radiative corrections \cite%
{Radiative}, systems with fermions \cite{fermions}, neutrinos \cite%
{neutrinos}, topological defects \cite{Defects}, topological phases \cite%
{Phases}, cosmic rays \cite{CosmicRay}, particle decays \cite{Iltan}, and
other relevant aspects \cite{Lehnert1}, \cite{General}. The SME has also
been used as a set-up to propose Lorentz-symmetry violation \cite{Tests} and
CPT- probing experiments \ \cite{CPT}, which have amounted to the imposition
of stringent bounds on the Lorentz-symmetry violating (LV) coefficients.

To take into account how this violation is implemented, in the fermion
sector of the SME, for example, there are two CPT-odd terms, $v_{\mu }%
\overline{\psi }\gamma ^{\mu }\psi ,b_{\mu }\overline{\psi }\gamma
_{5}\gamma ^{\mu }\psi $, where $v_{\mu },b_{\mu }$ are the LV backgrounds 
\cite{Hamilton}, \cite{Manojr}. A similar study has also been developed for
the case of a non-minimal coupling with the background, with new outcomes 
\cite{Nonmini}.\ Atomic and optical physics are other areas in which
Lorentz-symmetry violation has been intensively studied. Indeed, there are
several works examining Lorentz-violation in electromagnetic cavities and
optical systems \cite{Cavity}, \cite{Masers}, which contributed to establish
upper bounds on the LV coefficients.

The appearance of a more complex background, with the existence of an
anisotropic vacuum at Planck scale has drawn the attention of particle and
field theorists in recent years. Taking strings as the fundamental entities
at this level, the idea that Lorentz-symmetry and CPT invariance might be
spontaneously broken in some string theories became highlly estimulating 
\cite{Colladay, Samuel}. Very recently, the surprising result that
CPT-violating neutrino mass-squared difference would be an order of
magnitude less than the current uper bound on CPT-violation in the sector of
quarks and charged leptons has come to give more support to the proposal of
Lorentz-symmetry violation \cite{minos}, \cite{esposito}. Actually, the idea
of a CPT violation in the neutrino sector had been contemplated by Colladay,
Kostelecky and Mewes \cite{Colladay}.

Our main motivation to consider a supersymmetric scenario in connection with
the violation of Lorentz-symmetry is based on fact that we adopt the
viewpoint, according to \cite{Samuel}, that the breaking of Lorentz-symmetry
may be triggered by the vacuum condensation of a given tensor field.
Nevertheless, at this scale where Lorent-symmetry takes place, we may still
have supersymmetry (SUSY) or, even if SUSY had already been broken down by
some mechanism, supersymmetric partners are present and SUSY imprints are
not lost. So, by adopting this scenario, we argue that Lorentz-symmetry
violation, as originated at a more fundamental level in the string
framework, cannot be dissociated from SUSY. We do not have however elements
to decide whether SUSY breaking has occurred before Lorentz-symmetry
violation, though the breaking of the latter signals the violation of the
former. However, we would like to point out that, in the work of Ref. \cite%
{katz}, Katz and Shadmi, present an interesting discussion in the r\^{o}le
of vacuum expectation values that violate Lorentz symmetry as a possible to
realise SUSY breaking by $F$- and $D$- terms in the MSSM.

To implement effective Lorentz-symmetry-violating actions in\ supersymmetric
models, one may adopt a superspace formulation with Lorentz and, in some
cases, CPT invariances violated by a fixed background in Wess-Zumino-like
models \cite{berger}. In the gauge sector SUSY-preserving theories where
discussed in \cite{nib}. \ \ Also in the gauge sector, the proposed \cite%
{susy1} establishes a supersymmetric minimal extension for the
Chern-Simons-like term \cite{Jackiw} preserving the usual $(1+3)$%
-dimensional SUSY algebra. The breaking of SUSY will follow the very same
route to Lorentz-symmetry breaking: the statement that $v_{\mu }$ is a
constant shows thar SUSY is also broken by the fermionic SUSY \ partner of $%
v_{\mu }$.

Besides the CPT-odd terms, in the gauge sector, we have the CPT-even part,
which is represented by a tensor $K^{\mu \nu \alpha \beta }$ that presents
the same symmetries of the Riemann tensor, as well as an additional
double-traceless condition \cite{klin}. In this scenario, we present two
possibilities of constructing a supersymmetric version for the $K$-type
models.

So, we propose to carry out the supersymmetric extension to the bosonic
action below:

\begin{equation}
S=-\frac{1}{4}\int d^{4}x\;K_{\mu \nu \kappa \lambda }F_{{}}^{\mu \nu
}F^{\kappa \lambda }.  \label{acao1}
\end{equation}

The \textquotedblleft tensor\textquotedblright\ $K_{\mu \nu \kappa \lambda }$
it's CPT even, i. e., it does not violate the CPT-symmetry. Despite CPT
violation implies violation of Lorentz invariance \cite{greenberg}, the
reverse is not necesarily true. The action above is Lorentz-violanting in
the sense that the \textquotedblleft tensor\textquotedblright\ $K_{\mu \nu
\kappa \lambda }$ has a non-zero vacuum expectation value. That
\textquotedblleft tensor\textquotedblright\ presents the following
symmetries:

\begin{equation}  \label{anz1}
K_{\mu \nu \kappa \lambda }=K_{\left[ \mu \nu \right] \left[ \kappa \lambda %
\right] },\; K_{\mu \nu \kappa \lambda }=K_{\kappa \lambda \mu \nu },\; K{_{{%
\mu \nu }}}^{\mu \nu }=0,
\end{equation}

as usually appears in the literature, \ we can reduce the degrees of freedom
take into account the ansatz \cite{klin}:

\begin{equation}  \label{anz2}
K_{\mu \nu \kappa \lambda }=\frac{1}{2}\left( \eta _{\mu \kappa }\tilde{%
\kappa}_{\nu \lambda }-\eta _{\mu \lambda }\tilde{\kappa}_{\nu \kappa }+\eta
_{\nu \lambda }\tilde{\kappa}_{\mu \kappa }-\eta _{\nu \kappa }\tilde{\kappa}%
_{\mu \lambda }\right) ,
\end{equation}

\begin{equation}  \label{anz3}
\tilde{\kappa}_{\mu \nu }=\kappa \left( \xi _{\mu }\xi _{\nu }-\eta _{\mu
\nu }\xi ^{\alpha }\xi _{\alpha }/4 \right) ,
\end{equation}

\begin{equation}
\kappa =\frac{4}{3}\tilde{\kappa}^{\mu \nu }\xi _{\mu }\xi _{\nu },
\end{equation}

where $\tilde{\kappa}^{\mu \nu }$ is a traceless \textquotedblleft
tensor\textquotedblright . Using the ans\"{a}tze (\ref{anz2}), (\ref{anz3}),
in expression (\ref{acao1}), we obtain,

\begin{equation}
S=\frac{\kappa }{4}\int d^{4}x\left\{ \frac{1}{2}\xi _{\mu }\xi _{\nu }F{%
^{\mu }}_{\kappa }F^{\kappa \nu }+\frac{1}{8}\xi _{\lambda }\xi ^{\lambda
}F_{\mu \nu }F^{\mu \nu }\right\} .  \label{acao2}
\end{equation}

The supersymmetrization of the action (\ref{acao2}), instead of (\ref{acao1}%
), does not yield higher-spin components in the background dictated by SUSY.
To realize the supersymmetrization ,we have two possible routs: to achieve
supersymmetrisation of (\ref{acao1}) with $K_{\mu \nu \kappa \lambda }$
given by (\ref{anz2}), without the particular expression (\ref{anz3}) for $%
\tilde{\kappa}_{\mu \nu }$, we get that the background superfield must be
chiral. On the other hand, if we are to supersymetrise (\ref{acao1}) with $%
K_{\mu \nu \kappa \lambda }$ given by (\ref{anz2}) and $\tilde{\kappa}_{\mu
\nu }$ given by (\ref{anz3}), we conclude that a vector superfield must be
used to accommodate the background vector, $\xi ^{\lambda }$. The content of
partners is richer in this case than in the case of the chiral background.
The first route accommodate the chiral superfield as $\theta \sigma ^{\mu }%
\bar{\theta}\partial _{\mu }S$, while the second, is carried out by a vector
superfield accommodate this \textquotedblleft vector\textquotedblright\
background as $\theta \sigma ^{\mu }\bar{\theta}\xi _{\mu }$.

This work is organized as follows: In Section 2, we start with the model of
Lorentz breaking proposed as $K_{\mu \nu \kappa \lambda }$ by Kostelek\'{y}.
In this Section, we first study the possibility that the background tensor, $%
K_{\mu \nu \kappa \lambda }$, be originated from a background chiral
superfield, which imposes a special form - not yet discussed in the
literature - for the $K-$ tensor. Then in Section 3, we proceed \ by
investigating the possibility that $K_{\mu \nu \kappa \lambda }$ originates
from a (real) vector superfield, which exactly reproduces (\ref{acao1}) with
the $K-$ tensor defined by means of (\ref{anz2}), and (\ref{anz3}). Finally,
some Concluding Remarks are stated in Section 4.

\section{First Proposal: Lorentz-symmetry Breaking by a Chiral
Supermultiplet.}

Based on the work of Ref. \cite{susy1}, we take the idea that the background
vector could originate from a chiral supermultiplet, $\Omega $. As we shall
see, this imposes on $\xi _{\mu }$ the constraint $\xi _{\mu }=\partial
_{\mu }S$, where $S$ is a complex scalar. Indeed, as it will become clear at
the end of calculations, this choice of SUSY \ supermultiplet yields an
interesting form for $K_{\mu \nu \kappa \lambda }$, completely fixed by a
complex scalar field. We then show that the superspace action below shall
accomplish the task of yielding the component field extension of the action
of eq. (\ref{acao2}).\newline

Adopting covariant superspace-superfield formulation, we propose the
following minimal extension for: 
\begin{equation}
S=\kappa \int d^{4}xd^{2}\theta d^{2}\bar{\theta}\left\{ \left( D^{\alpha
}\Omega \right) W_{\alpha }(\overline{D}_{\dot{\alpha}}\overline{\Omega })%
\overline{W}^{\dot{\alpha}}+h.c.\right\} .  \label{superjac}
\end{equation}

The superfields $W_{a}$, $V$, $\Omega $ and the SUSY-covariant derivatives $%
\ D_{a}$, $\overline{D}_{\dot{a}}$ hold the definitions:

\begin{eqnarray}
D_{a} &=&\frac{\partial }{\partial \theta ^{a}}+i{\sigma }_{a\dot{a}}^{\mu }%
\bar{\theta}^{\dot{a}}\partial _{\mu }, \\
\overline{D}_{\dot{a}} &=&-\frac{\partial }{\partial \bar{\theta}^{\dot{a}}}%
-i{\theta }^{a}{\sigma }_{a\dot{a}}^{\mu }\partial _{\mu };
\end{eqnarray}%
from ${\overline{D}}_{\dot{b}}W_{a}\left( x,\theta ,\bar{\theta}\right) =0$%
,\ and $D^{a}W_{a}\left( x,\theta ,\bar{\theta}\right) =$ $\overline{D}_{%
\dot{a}}\overline{W}^{\dot{a}}\left( x,\theta ,\bar{\theta}\right) $, it
follows that 
\begin{equation}
W_{a}(x,\theta ,\bar{\theta})=-\frac{1}{4}\overline{D}^{2}D_{a}V.
\end{equation}%
Its $\theta $-expansion reads as below: 
\begin{eqnarray}
W_{a}(x,\theta ,\bar{\theta}) &=&\lambda _{a}\left( x\right) +i{\theta }^{b}{%
\ \sigma }_{b\dot{a}}^{\mu }\bar{\theta}^{\dot{a}}\partial _{\mu }\lambda
_{a}\left( x\right) -\frac{1}{4}{\bar{\theta}}^{2}\theta ^{2}\square \lambda
_{a}\left( x\right) +  \notag \\
&+&2\theta _{a}D\left( x\right) -i{\theta }^{2}\bar{\theta}^{\dot{a}}{\sigma 
}_{a\dot{a}}^{\mu }\partial _{\mu }D\left( x\right) +{{{\sigma }^{\mu \nu }}%
_{a}}^{b}\theta _{b}F_{\mu \nu }\left( x\right)  \notag \\
&-&\frac{i}{2}{{{\sigma }^{\mu \nu }}_{a}}^{b}{\sigma }_{b\dot{a}}^{\alpha
}\theta ^{2}\bar{\theta}^{\dot{a}}\partial _{\alpha }F_{\mu \nu }\left(
x\right) -i\sigma _{a\dot{a}}^{\mu }\partial _{\mu }\text{ }\bar{\lambda}^{%
\dot{a}}\left( x\right) \theta ^{2},
\end{eqnarray}

and $V=V^{\dagger }$. The Wess-Zumino gauge choice is taken as usually done: 
\begin{equation}
\text{ }V_{WZ}=\theta \sigma ^{\mu }\bar{\theta}A_{\mu }(x)+\theta ^{2}\bar{%
\theta}\overline{\lambda }\left( x\right) +\bar{\theta}^{2}\theta \lambda
(x)+\theta ^{2}\bar{\theta}^{2}D,  \label{vess}
\end{equation}%
so the\ action (\ref{superjac}) is gauge-invariant. The background
superfield is so chosen to be a chiral one. Such a constraint restricts the
maximum spin component of the background to be an $s=$ $\frac{1}{2}$
component-field, showing up as a SUSY-partner for a spinless dimensionless
scalar field. Also, one should notice that $\Omega $ has dimension of mass
to $-1$. The superfield expansion for $\Omega $ then reads:

\begin{eqnarray}
\overline{D}_{\dot{a}}\Omega \left( x,\theta ,\bar{\theta}\right) &=&0,\,%
\text{ }  \notag \\
\text{\ consequently }\,\Omega \left( x,\theta ,\bar{\theta}\right) \text{ }
&=&S\left( x\right) +\sqrt{2}\theta \zeta \left( x\right) +i\theta {\sigma }%
^{\mu }\bar{\theta}\partial _{\mu }S\left( x\right) +  \notag \\
&+&\theta ^{2}G\left( x\right) +\frac{i}{\sqrt{2}}{\theta }^{2}\bar{\theta}%
\bar{\sigma}^{\mu }\partial _{\mu }\zeta \left( x\right) -\frac{1}{4}{\bar{%
\theta}}^{2}{\theta }^{2}\square S\left( x\right) .
\end{eqnarray}

With its SUSY transformations given by

\begin{eqnarray}
\delta S &=&\sqrt{2}\varepsilon ^{\alpha }\zeta _{\alpha },  \notag \\
\delta \zeta _{\alpha } &=&\sqrt{2}G\varepsilon _{\alpha }+i\sqrt{2}{\sigma }%
_{\alpha \dot{\alpha}}^{\mu }\text{ }\bar{\varepsilon}^{\dot{\alpha}%
}\partial _{\mu }S,  \notag \\
\delta G &=&i\sqrt{2}\bar{\varepsilon}_{\dot{\alpha}}^{{}}{\bar{\sigma}}%
_{{}}^{\mu \dot{\alpha}\alpha }\text{ }\partial _{\mu }\zeta _{\alpha }.
\label{susy1}
\end{eqnarray}

Notice that, if we wish to have $\partial _{\mu }S$ constant (as we shall
get in the sequel, a constant $\partial _{\mu }S$ give us a constant $K_{\mu
\nu \kappa \lambda }$), so that $S$ depends lenearly on $x^{\mu }$, SUSY is
also broken by the backgroud, as the expression for $\delta \zeta $ shows:
if we apply a SUSY transformation on an $\Omega $-superfield for which $%
S\neq 0$, $\zeta =0$, $G=0$, then $\delta \zeta \neq 0$ whenever $\partial
_{\mu }S\neq 0$. So, to realise the breaking of Lorentz-symmetry in terms of
an $\Omega $ with a constant $\partial _{\mu }S$, then SUSY is not an
invariance of the background $\Omega $. However, as it stands, we have an
effective model which may descend from a more fundamental theory in which
SUSY might be spontaneously broken. At this stage, we have no commitment
with any specific mechanism for SUSY breaking.\newline

The component-wise counterpart for the action (\ref{superjac}) is as follows%
\footnote{%
the ref. \cite{haber} has been used in our calculations}:

\begin{equation}
{S}=\kappa \int d^{4}x\,d^{2}\theta \,d^{2}\bar{\theta}\left\{ \left(
D^{\alpha }\Omega \right) W_{\alpha }(\overline{D}_{\dot{\alpha}}{\overline{%
\Omega }})\overline{W}^{\dot{\alpha}}+h.c.\right\}
=S_{boson}+S_{fermion}+S_{coupled},  \label{15}
\end{equation}

\begin{eqnarray}
S_{boson} &=&\kappa \int d^{4}x\Bigg\{-16\left[ \frac{1}{4}\partial
_{\lambda }S\partial _{\mu }S^{\ast }({F^{\mu }}_{\kappa }F^{\kappa \lambda
}+{F^{\lambda }}_{\kappa }F^{\kappa \mu })+\frac{1}{8}\partial _{\lambda }S{%
\partial }{}^{\lambda }S^{\ast }F_{\mu \nu }F^{\mu \nu }\right] +  \notag \\
&+&2D\varepsilon ^{\lambda \mu \tau \rho }\partial _{\lambda }S\partial
_{\mu }S^{\ast }F_{\tau \rho }-2i\partial _{\lambda }S\partial _{\mu
}S^{\ast }\varepsilon ^{\lambda \tau \rho \kappa }F_{\tau \rho }{F^{\mu }}%
_{\kappa }+8iD\partial _{\lambda }S\partial _{\mu }S^{\ast }F^{\lambda \mu }+
\notag \\
&&-2D\partial _{\lambda }S\partial _{\mu }S^{\ast }\varepsilon ^{\lambda \mu
\nu \kappa }F_{\nu \kappa }+8D^{2}\partial _{\mu }S\partial ^{\mu }S^{\ast
}+16D^{2}\left\vert G\right\vert ^{2}+h.c.\Bigg\},
\end{eqnarray}

\begin{eqnarray}
S_{fermion} &=&\kappa \int d^{4}x\Bigg\{\frac{1}{2}{\partial }_{\lambda
}\zeta \sigma ^{\mu }\partial _{\mu }\bar{\zeta}\lambda \sigma ^{\lambda }%
\bar{\lambda}+\frac{1}{2}{\partial }_{\lambda }\zeta \sigma ^{\mu }\bar{%
\lambda}\lambda \sigma ^{\lambda }\partial _{\mu }\bar{\zeta}+2{\partial }%
_{\mu }\zeta \partial ^{\mu }\lambda \bar{\zeta}\bar{\lambda}+  \notag \\
&-&\frac{1}{2}{\partial }_{\lambda }\zeta {\sigma }_{{}}^{\lambda }{\partial 
}_{\mu }\bar{\zeta}\lambda {\sigma ^{\mu }}\bar{\lambda}-2{\partial }%
_{\lambda }\zeta {\sigma _{{}}^{\lambda }\bar{\sigma}{^{\mu }}\partial }%
_{\mu }\lambda \bar{\zeta}\bar{\lambda}-\frac{1}{2}\lambda \sigma ^{\lambda }%
{\bar{\sigma}{^{\mu }\partial }}_{\lambda }\zeta \bar{\zeta}\partial _{\mu }%
\bar{\lambda}+  \notag \\
&-&\frac{1}{2}\zeta \lambda \bar{\zeta}\square \bar{\lambda}-\zeta \lambda
\partial _{\mu }\bar{\zeta}\bar{\sigma}{^{\mu }\sigma ^{\tau }}\partial
_{\tau }\bar{\lambda}+\frac{1}{2}\zeta \lambda \partial _{\mu }\bar{\lambda}%
\bar{\sigma}{^{\nu }\sigma }^{\mu }\partial _{\nu }\bar{\zeta}+  \notag \\
&+&\frac{1}{2}\partial _{\mu }\zeta {\sigma ^{\mu }}\bar{\sigma}{^{\nu
}\lambda }\bar{\zeta}{\partial }_{\nu }\bar{\lambda}-\frac{1}{2\sqrt{2}}%
\zeta \square \lambda \bar{\zeta}\bar{\lambda}-\frac{1}{2}\zeta \partial
_{\nu }\lambda \partial _{\mu }\bar{\lambda}\bar{\sigma}{^{\mu }\sigma ^{\nu
}}\bar{\lambda}+  \notag \\
&-&\frac{1}{2}\zeta \partial _{\nu }\lambda \partial _{\mu }\bar{\zeta}\bar{%
\sigma}{^{\nu }\sigma ^{\mu }}\bar{\lambda}+\zeta \partial _{\mu }\lambda 
\bar{\zeta}\partial ^{\mu }\bar{\lambda}-2\zeta {\sigma }^{\mu }\partial
_{\mu }\bar{\lambda}\bar{\zeta}\bar{\sigma}{^{\nu }}\partial _{\nu }\lambda
+h.c.\Bigg\},
\end{eqnarray}

\begin{eqnarray}
S_{coupled} &=&\kappa \int d^{4}x\Bigg\{-4iD\zeta {\sigma }^{\mu }\partial
_{\mu }\bar{\zeta}-2\sqrt{2}iDG^{\ast }\zeta {\sigma }^{\mu }\partial _{\mu }%
\bar{\lambda}+2\sqrt{2}D{\partial }_{\nu }\lambda {\sigma }^{\nu }\bar{\sigma%
}{^{\mu }\zeta }\partial _{\mu }S^{\ast }+  \notag \\
&+&2D\zeta {\sigma ^{\nu }\partial }_{\mu }\bar{\zeta}F_{\nu }{}^{\mu
}+iD\varepsilon ^{\tau \rho \mu \alpha }\zeta \sigma _{\alpha }\partial
_{\mu }\bar{\zeta}F_{\tau \rho }+\sqrt{2}G^{\ast }\zeta {\sigma }^{\mu
}\partial _{\nu }\bar{\lambda}F_{\mu }{}^{\nu }+  \notag \\
&+&\frac{i}{\sqrt{2}}G^{\ast }\varepsilon ^{\tau \rho \mu \alpha }\zeta
\sigma _{\alpha }\partial _{\mu }\bar{\lambda}F_{\tau \rho }+\sqrt{2}i\zeta {%
\sigma }^{\tau }\bar{\sigma}{^{\nu }}\partial _{\nu }\lambda \partial _{\mu
}S^{\ast }F_{\tau }{}^{\mu }+  \notag \\
&-&\frac{1}{\sqrt{2}}\varepsilon ^{\tau \rho \mu \alpha }\zeta \sigma
_{\alpha }\bar{\sigma}{^{\nu }}\partial _{\mu }S^{\ast }\partial _{\nu
}\lambda F_{\tau \rho }-4\sqrt{2}iG^{\ast }D\zeta {\sigma }^{\mu }\partial
_{\mu }\bar{\lambda}+  \notag \\
&+&2\sqrt{2}D\zeta \partial _{\mu }\lambda \partial ^{\mu }S^{\ast }-\frac{i%
}{\sqrt{2}}\varepsilon ^{\mu \nu \kappa \tau }\zeta \partial _{\tau }\lambda
\partial _{\mu }S^{\ast }F_{\nu \kappa }+\frac{1}{2\sqrt{2}}\varepsilon
^{\mu \nu \kappa \tau }\zeta \partial _{\tau }\lambda \partial _{\mu
}S^{\ast }F_{\nu \kappa }+  \notag \\
&-&4iD^{2}\bar{\zeta}\bar{\sigma}{^{\mu }}\partial _{\mu }\zeta -2D\bar{\zeta%
}\bar{\sigma}{^{\nu }}\partial _{\mu }\zeta F_{\nu }{}^{\mu }+iD\varepsilon
^{\nu \kappa \mu \alpha }\bar{\zeta}\bar{\sigma}_{\alpha }\partial _{\mu
}\zeta F_{\nu \kappa }+  \notag \\
&+&2\sqrt{2}iDG^{\ast }\partial _{\mu }\zeta {\sigma }^{\mu }\bar{\lambda}%
+2D\partial _{\mu }\zeta {\sigma ^{\tau }}\bar{\zeta}F_{\tau }{}^{\mu
}+iD\varepsilon ^{\tau \rho \mu \alpha }\partial _{\mu }\zeta \sigma
_{\alpha }\bar{\zeta}F_{\tau \rho }+  \notag \\
&+&2\partial _{\mu }\zeta {\sigma }^{\tau }\bar{\sigma}{^{\nu \kappa }}\bar{%
\zeta}F_{\nu \kappa }F_{\tau }{}^{\mu }+i\varepsilon ^{\tau \rho \mu \alpha
}\partial _{\mu }\zeta \sigma _{\alpha }\bar{\sigma}{^{\nu \kappa }}\bar{%
\zeta}F_{\tau \rho }{}F_{\nu \kappa }+\sqrt{2}\partial _{\mu }\zeta {\sigma }%
^{\tau }\bar{\lambda}F_{\tau }{}^{\mu }+  \notag \\
&+&\frac{i}{\sqrt{2}}\varepsilon ^{\tau \rho \mu \alpha }\partial _{\mu
}\zeta \sigma _{\alpha }\bar{\lambda}F_{\tau \rho }-2\sqrt{2}iGD\lambda {%
\sigma ^{\mu }}\partial _{\mu }\bar{\zeta}+2i\left\vert G\right\vert
^{2}\lambda {\sigma ^{\mu }}\partial _{\mu }\bar{\lambda}+  \notag \\
&-&2i\partial _{\nu }\lambda \sigma ^{\nu }\bar{\sigma}{^{\mu }\lambda
\partial }_{\mu }S^{\ast }+4\sqrt{2}iGD\partial _{\mu }\lambda \sigma ^{\mu }%
\bar{\zeta}-2\sqrt{2}iGD\bar{\zeta}\bar{\sigma}{^{\mu }\partial }_{\mu }{%
\lambda }\bar{\zeta}+  \notag \\
&-&\sqrt{2}G\bar{\zeta}\bar{\sigma}{^{\mu }\partial }_{\tau }{\lambda }%
F_{\mu }{}^{\tau }+\frac{i}{\sqrt{2}}G\varepsilon ^{\mu \nu \tau \alpha }%
\bar{\zeta}\bar{\sigma}_{\alpha }\partial _{\tau }\lambda F_{\mu \nu
}-2i\left\vert G\right\vert ^{2}\bar{\lambda}\bar{\sigma}{^{\mu }}\partial
_{\mu }\lambda +  \notag \\
&+&2\sqrt{2}D\partial _{\mu }S\bar{\zeta}\partial ^{\mu }\bar{\lambda}+\sqrt{%
2}i{\partial }_{\mu }\left( \bar{\lambda}\bar{\zeta}\right) \partial
_{\lambda }SF^{\lambda \mu }-\frac{1}{\sqrt{2}}\varepsilon ^{\mu \lambda
\tau \rho }{\partial }_{\mu }\left( \bar{\lambda}\bar{\zeta}\right) \partial
_{\lambda }SF_{\tau \rho }+  \notag \\
&-&2\sqrt{2}D\bar{\lambda}{\partial }_{\mu }\bar{\zeta}\partial ^{\mu }S+2%
\sqrt{2}D\bar{\zeta}\bar{\sigma}{^{\mu }\sigma ^{\nu }\partial }_{\nu }\bar{%
\lambda}\partial _{\mu }S-\sqrt{2}i\bar{\zeta}\bar{\sigma}{^{\nu }\sigma
^{\mu }\partial }_{\mu }\bar{\lambda}F_{\nu \lambda }{\partial }{}^{\lambda
}S+  \notag \\
&-&\frac{1}{\sqrt{2}}\varepsilon ^{\nu \kappa \lambda \alpha }\bar{\zeta}%
\bar{\sigma}_{\alpha }{\sigma ^{\mu }\partial }_{\mu }\bar{\lambda}F_{\nu
\kappa }\partial _{\lambda }S+2G^{\ast }\bar{\lambda}\bar{\sigma}{^{\mu
}\sigma }^{\nu }\partial _{\nu }\bar{\lambda}\partial _{\mu }S-2\sqrt{2}%
\partial _{\mu }S\partial ^{\mu }D\bar{\zeta}\bar{\lambda}+  \notag \\
&+&\sqrt{2}D\partial _{\nu }\zeta {\sigma ^{\nu }\bar{\sigma}{^{\mu }}%
\lambda }\partial _{\mu }S^{\ast }-\frac{i}{\sqrt{2}}\bar{\sigma}{^{\lambda
}\partial }_{\lambda }\zeta {\lambda }{\sigma }^{\nu }{\partial }_{\mu
}S^{\ast }F_{\nu }{}^{\mu }+\frac{1}{2\sqrt{2}}\varepsilon ^{\mu \nu \kappa
\alpha }\lambda \sigma _{\alpha }\bar{\sigma}{^{\lambda }\partial }_{\lambda
}\zeta {\partial }_{\mu }S^{\ast }F_{\nu \kappa }+  \notag \\
&-&\frac{1}{2}{\partial }_{\lambda }\zeta {\sigma }_{{}}^{\lambda }\bar{%
\lambda}\lambda {\sigma ^{\mu }\partial }_{\mu }\bar{\zeta}+\sqrt{2}D\lambda
\sigma ^{\lambda }{\bar{\sigma}{^{\mu }\partial }}_{\lambda }\zeta \partial
_{\mu }S^{\ast }-\frac{i}{\sqrt{2}}{\partial }_{\lambda }\zeta \sigma ^{\nu }%
{\bar{\sigma}{^{\lambda }\lambda }}\partial _{\mu }S^{\ast }F_{\nu }{}^{\mu
}+  \notag \\
&-&\frac{1}{2\sqrt{2}}\varepsilon ^{\mu \nu \kappa \alpha }{\partial }%
_{\lambda }\zeta \sigma _{\alpha }\bar{\sigma}{^{\lambda }\lambda }\partial
_{\mu }S^{\ast }F_{\nu \kappa }+h.c.\Bigg\}.  \label{smix1}
\end{eqnarray}%
\ We notice that, in trying to supersymmetrise the $\xi _{\mu }\xi _{\nu
}F_{\kappa }^{\mu }F^{\kappa \nu }$-term, we automatically get\ the
supersymmetrisation of the term $\xi _{\lambda }\xi ^{\lambda }F^{2}$. This
is not a simple coincidence, but it can naturally be expected from an
analysis of the irreducible representations of \ $SO(1,3)$. We conclude
that, in the case the background superfield is chiral, we are able to write
down the supersymmetric version of (\ref{acao1}) with $K_{\mu \nu \kappa
\lambda }$ given by \cite{klin}:

\begin{equation}
K_{\mu \nu \kappa \lambda }=-16\left(\eta _{\substack{ k\mu  \\ }}\tilde{%
\kappa}_{\nu \lambda }-\eta _{\substack{ \mu \lambda  \\ }}\tilde{\kappa}%
_{\nu \kappa }-\eta _{\substack{ \nu \lambda  \\ }}\tilde{\kappa}_{\mu
\kappa }-\eta _{\substack{ k\nu  \\ }}\tilde{\kappa}_{\mu \lambda }\right),
\end{equation}

with,

\begin{equation}
\tilde{\kappa}_{\mu \nu }=\kappa \left\{ \left( \frac{\partial _{\mu
}S\partial _{\nu }S^{\ast }+\partial _{\nu }S\partial _{\mu }S^{\ast }}{2}%
\right) -\partial _{\lambda }S\partial ^{\lambda }S^{\ast }\eta _{\substack{ %
\mu \nu  \\ }}/4\right\} ,
\end{equation}

and

\begin{equation}
K{^{\mu \nu }}_{\mu \nu }=0.
\end{equation}

Concluding, this special form for $\tilde{\kappa}$ is a natural consequence
of the assumption that a chiral superfield carries the background that
breaks Lorentz-symmetry.\newline
It is worthy to notice that $\tilde{\kappa}$, and consequently $K$, depend
exclusively on the scalar component $S$. No $\zeta $-condensate and no $G$%
-dependence appear in the expression for the $K$-tensor. Since the action (%
\ref{15}) is quadratic in $\Omega $, one might in principle expect that
tensor bilinears in $\zeta $ could show up as contributions to $K_{\mu \nu
\kappa \lambda }$. However, the $K$-tensor depends exclusively on the
gradient of $S$. It becomes clear that a constant $\partial _{\mu }S$
ensures the constancy of $K_{\mu \nu \kappa \lambda }$, as we had already
anticipated. In a particular case, where the background fields $\partial
_{\mu }S\neq 0$, $\zeta =0,$ and $G=0$, we have that the auxiliar field, $D$%
, is given by ( we suppose the supersymmetric version of the $F_{\mu \nu
}^{2}$-term is added up),

\begin{equation}
D=-\frac{8\kappa i\partial _{\mu }S\partial _{\nu }S^{\ast }F^{\mu \nu }}{%
16+32\kappa \partial _{\mu }S\partial ^{\mu }S^{\ast }}.  \label{22}
\end{equation}

It is interesting to comment that, by virtue of (\ref{22}), which is valid
in the conditions above for the background, the breaking of
Lorentz-symmetry, fixes the auxiliary field to be non-trivial, even, if the
gauge potential superfield $(V)$ is not coupled to matter. The backgound, as
(\ref{22}) shows, determines $D$ by the gauge field strenght, $F_{\mu \nu }$%
. However, if supersymmetric matter happens to be minimally coupled to the
gauge field, then (\ref{22}) indicates that charged scalar particles
(selectrons, for example) may acquire a magnetic dipole moment given in
terms of the vector $\vec{\mu}=\vec{v}\times \vec{v}^{\ast }$, where $\vec{v}%
\equiv \vec{\nabla}S$. This investigation is being pursued and we shall soon
report on it \cite{fut}.

\section{Second Proposal: Lorentz-symmetry Breaking from a Vector
Supermultiplet.}

\bigskip Adopting covariant superspace-superfield formulation, we propose
the following minimal extension of (\ref{acao2}) for: 
\begin{equation}
S=\kappa \int d^{4}xd^{2}\theta d^{2}\bar{\theta}\left\{ \left( D^{\alpha
}\Xi \right) W_{\alpha }(\overline{D}_{\dot{\alpha}}\Xi )\overline{W}^{\dot{%
\alpha}}+h.c.\right\} .
\end{equation}

$W_{\alpha }$ is the superfield strenght of the gauge supermultiplet $V_{WZ}$%
, as given in eq. (\ref{acao1}). $\Xi $ is the so-called vector superfield,
whose $\theta -$ expansion is as follows:

\begin{equation}
\Xi \left( \theta ,\bar{\theta}\right) =C+\theta \chi +\bar{\theta}\bar{\chi}%
+\theta ^{2}M+\bar{\theta}^{2}{M}^{\ast }+\theta \sigma ^{\mu }\bar{\theta}%
\xi _{\mu }+\theta ^{2}\bar{\theta}\bar{\psi}+{\bar{\theta}}^{2}{\theta }{%
\psi }+\theta ^{2}\bar{\theta}^{2}B.
\end{equation}

In the special case of constant background component fields, the SUSY
transformations simplify and acquire the form below:

\begin{eqnarray}
\delta C &=&\varepsilon ^{\alpha }\chi _{\alpha }+\bar{\varepsilon}_{\dot{%
\alpha}}\bar{\chi}^{\dot{\alpha}},  \notag \\
\delta \chi _{\alpha } &=&2M\varepsilon _{\alpha }+{\sigma }_{\alpha \dot{%
\alpha}}^{\mu }\bar{\varepsilon}^{\dot{\alpha}}\xi _{\mu },  \notag \\
\delta \bar{\chi}^{\dot{\alpha}} &=&-\varepsilon ^{\alpha }{\sigma }_{\alpha 
\dot{\beta}}^{\mu }\varepsilon ^{\dot{\beta}\dot{a}}\xi _{\mu }+2M^{\ast }%
\bar{\varepsilon}^{\dot{\alpha}},  \notag \\
\delta M &=&\bar{\varepsilon}_{\dot{\alpha}}\bar{\psi}^{\dot{\alpha}}, 
\notag \\
\delta M^{\ast } &=&\varepsilon _{{}}^{\alpha }\chi _{\alpha },  \notag \\
\delta \xi ^{\mu } &=&\varepsilon ^{\alpha }{\sigma }_{\alpha \dot{\alpha}%
}^{\mu }\bar{\psi}^{\dot{\alpha}}-\bar{\varepsilon}_{\dot{\alpha}}{\bar{%
\sigma}}^{\mu \dot{\alpha}\alpha }\psi _{\alpha },  \notag \\
\delta \bar{\psi}^{\dot{\alpha}} &=&2\bar{\varepsilon}^{\dot{\alpha}}B, 
\notag \\
\delta \psi _{\beta } &=&2\varepsilon _{\beta }B,  \notag \\
\delta B &=&0.  \label{susy2}
\end{eqnarray}

In the general case, $B$ transforms as a total derivative; only for a
constant $\psi $-background, we get $\delta B=0.$ Notice that we are taking
the full $\theta $-expansion for $\Xi .$ Nothing like a Wess-Zumino gauge
can be taken, for $\Xi $ is a fixed background and is not a gauge
super-potential.

The superaction in component fields is given by,

\begin{equation}
{S}=\kappa \int d^{4}x\,d^{2}\theta \,d^{2}\bar{\theta}\Big\{(D^{\alpha }\Xi
)W_{\alpha }({\overline{D}}_{\dot{\alpha}}\Xi ){\overline{W}}^{\dot{\alpha}%
}+h.c.\Big\}=S_{boson}+S_{fermion}+S_{coupled};  \label{26}
\end{equation}

\begin{eqnarray}
S_{boson} &=&\kappa \int d^{4}x\,\Bigg\{2\left( \frac{1}{2}\xi _{\lambda
}\xi _{\mu }F^{\mu }{}_{\kappa }F^{\kappa \lambda }+\frac{1}{8}\xi _{\lambda
}\xi ^{\lambda }F_{\mu \nu }F^{\mu \nu }\right) +  \notag \\
&+&\frac{i}{4}\xi _{\lambda }\xi _{\mu }\varepsilon ^{\lambda \tau \rho
\kappa }F_{\tau \rho }F^{\mu }{}_{\kappa }-2D^{2}\xi _{\lambda }\xi
^{\lambda }+8{\left\vert M\right\vert }^{2}D^{2}+h.c.\Bigg\},
\end{eqnarray}

where we can readily identify the action (\ref{acao1}) with $K_{\mu \nu
\kappa \lambda }$ given as in (\ref{anz2}), and (\ref{anz3}).

\begin{eqnarray}
S_{fermion} &=&\kappa \int d^{4}x\,\Bigg\{-i\psi \sigma ^{\mu }\partial
_{\mu }\bar{\lambda}\bar{\chi}\bar{\lambda}+\psi \lambda \bar{\psi}\bar{%
\lambda}+\frac{i}{2}\partial _{\mu }\lambda \sigma ^{\mu }\bar{\psi}\bar{\chi%
}\bar{\lambda}+\psi \lambda \partial _{\mu }\lambda \sigma ^{\mu }\bar{\chi}+
\notag \\
&-&\frac{i}{2}\lambda \sigma ^{\mu }\bar{\psi}\bar{\chi}\partial _{\mu }\bar{%
\lambda}+\lambda ^{2}\bar{\psi}\bar{\lambda}+i\chi \lambda \partial _{\mu
}\lambda \sigma ^{\mu }\overline{\psi }-i\chi \lambda \psi {\sigma }^{\mu
}\partial _{\mu }\bar{\lambda}+h.c.\Bigg\},
\end{eqnarray}

\begin{eqnarray}
S_{coupled} &=&\kappa \int d^{4}x\,\Bigg\{-\frac{3}{4}i\chi \lambda \xi
_{\mu }\partial ^{\mu }D-\frac{1}{2}\chi \lambda \xi _{\mu }\partial _{\nu
}F^{\nu }{}_{\mu }+  \notag \\
&-&\frac{i}{2}\chi \lambda \varepsilon ^{\psi \nu \kappa \mu }\xi _{\mu
}\partial _{\psi }F_{\nu \kappa }+4\chi \lambda BD+D\chi \sigma ^{\mu }\bar{%
\chi}\partial _{\nu }F^{\nu }{}_{\mu }+  \notag \\
&+&\frac{i}{4}D\varepsilon ^{\psi \nu \kappa \alpha }\chi \sigma _{\alpha }%
\bar{\chi}\partial _{\psi }F_{\nu \kappa }-6iDM^{\ast }\chi \sigma ^{\mu
}\partial _{\mu }\bar{\lambda}+iD\partial _{\nu }\lambda \sigma ^{\nu }\bar{%
\sigma}^{\mu }\chi \xi _{\mu }+  \notag \\
&-&4D^{2}\chi \psi -\frac{i}{2}\chi \sigma ^{\tau }\bar{\chi}F_{\tau \rho
}\partial _{\nu }F^{\nu \rho }+\frac{1}{4}\varepsilon ^{\tau \rho \kappa
\alpha }\chi \sigma _{\alpha }\bar{\chi}F_{\tau \rho }\partial _{\nu
}F{}^{\nu }{}_{\kappa }+  \notag \\
&+&\frac{1}{4}\varepsilon ^{\lambda \nu \kappa \rho }\chi \sigma ^{\tau }%
\bar{\chi}F_{\tau \rho }\partial _{\lambda }F_{\nu \kappa }+\frac{i}{8}%
\varepsilon ^{\psi \nu \kappa \alpha }\varepsilon {}^{\tau \rho }{}_{\alpha
\beta }\chi \sigma ^{\beta }\bar{\chi}F_{\tau \rho }\partial _{\psi }F_{\nu
\kappa }+  \notag \\
&+&M^{\ast }\chi \sigma ^{\tau }\partial _{\nu }\bar{\lambda}F_{\tau
}{}^{\nu }+\frac{i}{2}M^{\ast }\varepsilon ^{\tau \rho \mu \alpha }\chi {%
\sigma }_{\alpha }\partial _{\mu }\bar{\lambda}F_{\tau \rho }+\frac{1}{2}%
\chi \sigma ^{\tau }\bar{\sigma}^{\nu }\partial _{\nu }\lambda \xi _{\mu
}F_{\tau }{}^{\mu }+  \notag \\
&+&\frac{i}{4}\varepsilon ^{\tau \rho \mu \alpha }\sigma _{\alpha }\bar{%
\sigma}^{\nu }\partial _{\nu }\lambda \xi _{\mu }F_{\tau \rho }-2D\chi
\sigma ^{\tau \rho }\lambda F_{\rho \tau }+\frac{1}{2}\chi \partial _{\mu
}\lambda \bar{\chi}\partial ^{\mu }\bar{\lambda}+  \notag \\
&+&i\chi \partial _{\mu }\lambda \xi ^{\mu }-\bar{\chi}\bar{\sigma}^{\nu
}\chi \partial _{\mu }DF_{\nu }{}^{\mu }+\frac{i}{2}\varepsilon ^{\nu \kappa
\mu \alpha }\bar{\chi}\bar{\sigma}_{\alpha }\chi \partial _{\mu }DF_{\nu
\kappa }+  \notag \\
&+&iMD\partial _{\mu }\lambda \sigma ^{\mu }\bar{\chi}+\frac{i}{2}M\lambda
\sigma ^{\kappa }\bar{\chi}\partial _{\mu }F^{\mu }{}_{\kappa }-\frac{1}{4}%
M\varepsilon ^{\mu \nu \kappa \alpha }\lambda \sigma _{\alpha }\bar{\chi}%
\partial _{\mu }F_{\nu \kappa }+  \notag \\
&-&i{\left\vert M\right\vert }^{2}\lambda \sigma ^{\mu }\partial _{\mu }\bar{%
\lambda}+iM\partial _{\nu }\lambda \sigma ^{\nu }\bar{\sigma}^{\mu }\lambda {%
\xi }_{\mu }-4MD\lambda ^{2}+  \notag \\
&-&\frac{1}{2}\bar{\chi}\bar{\sigma}^{\nu }\partial _{\mu }\lambda F_{\nu
}{}^{\mu }+\frac{i}{4}\varepsilon ^{\nu \kappa \mu \alpha }\bar{\chi}\bar{%
\sigma}_{\alpha }\partial _{\mu }\lambda F_{\nu \kappa }+\frac{i}{2}M^{\ast
}\partial _{\mu }\lambda \sigma ^{\mu }\bar{\lambda}+  \notag \\
&+&i\lambda \sigma ^{\mu }\partial _{\nu }\bar{\lambda}\xi ^{\nu }\xi _{\mu
}-\frac{i}{2}\lambda \sigma ^{\mu }\partial _{\mu }\bar{\lambda}\xi
_{\lambda }\xi ^{\lambda }-\frac{1}{2}\varepsilon ^{\lambda \nu \mu \alpha
}\lambda \sigma _{\alpha }\partial _{\nu }\bar{\lambda}\xi _{\lambda }\xi
_{\mu }+  \notag \\
&+&\frac{1}{2}\varepsilon ^{\nu \kappa \lambda \alpha }\lambda \sigma
_{\alpha }\bar{\psi}\xi _{\lambda }F_{\nu \kappa }-B\lambda \sigma ^{\mu }%
\bar{\lambda}\xi _{\mu }+\bar{\chi}\partial _{\nu }\bar{\lambda}\xi _{\mu
}F^{\mu \nu }+  \notag \\
&+&\frac{i}{2}\varepsilon ^{\tau \rho \nu \lambda }\bar{\chi}\partial _{\nu }%
\bar{\lambda}\xi _{\lambda }F_{\tau \rho }-\frac{i}{2}\bar{\lambda}\bar{%
\sigma}^{\rho }\lambda \xi _{\lambda }F^{\lambda }{}_{\rho }-\frac{1}{4}%
\varepsilon ^{\lambda \tau \rho \alpha }\bar{\lambda}\bar{\sigma}_{\alpha
}\lambda \xi _{\lambda }F_{\tau \rho }+  \notag \\
&-&iD\partial _{\mu }\bar{\lambda}\bar{\sigma}^{\mu }\sigma ^{\lambda }\bar{%
\chi}\xi _{\lambda }+\frac{1}{2}\partial _{\mu }\bar{\lambda}\bar{\sigma}%
^{\mu }\sigma ^{\kappa }\bar{\chi}\xi _{\lambda }F^{\lambda }{}_{\kappa }+%
\frac{i}{4}\varepsilon ^{\lambda \nu \kappa \alpha }\partial _{\mu }\bar{%
\lambda}\bar{\sigma}^{\mu }\sigma _{\alpha }\bar{\chi}\xi _{\lambda }F_{\nu
\kappa }+  \notag \\
&-&iM^{\ast }\partial _{\mu }\bar{\lambda}\bar{\sigma}^{\mu }\sigma
^{\lambda }\bar{\lambda}\xi _{\lambda }-\frac{1}{2}\bar{\chi}\bar{\lambda}%
\xi _{\lambda }\partial _{\nu }F^{\lambda \nu }-\frac{i}{4}\bar{\chi}\bar{%
\lambda}\varepsilon ^{\tau \rho \nu \lambda }\xi _{\lambda }\partial _{\nu
}F_{\tau \rho }+  \notag \\
&-&\frac{1}{2}D^{2}\bar{\chi}\bar{\psi}+D\bar{\chi}\bar{\sigma}^{\nu \kappa }%
\bar{\psi}F_{\nu \kappa }-2M^{\ast }D\bar{\psi}\bar{\lambda}-\frac{i}{2}\psi
\sigma ^{\tau }\bar{\lambda}\xi _{\mu }F_{\tau }{}^{\mu }+  \notag \\
&+&\frac{1}{4}\varepsilon ^{\tau \rho \mu \alpha }\psi \sigma _{\alpha }\bar{%
\lambda}\xi _{\mu }F_{\tau \rho }-B\lambda \sigma ^{\mu }\bar{\lambda}\xi
_{\mu }+4BD\bar{\chi}\bar{\lambda}+h.c.\Bigg\}.  \label{smix2}
\end{eqnarray}

The terms in the superactions (\ref{15}) and (\ref{26}) \ have the same form
with the factor $\left( D^{\alpha }\Omega \right) W_{\alpha }$ replaced by $%
\left( D^{\alpha }\Xi \right) W_{\alpha }$. Then, we obtain the right
combination, i. e., $\frac{1}{2}\xi _{\mu }\xi _{\nu }F{^{\mu }}_{\kappa
}F^{\kappa \nu }+\frac{1}{8}\xi _{\lambda }\xi ^{\lambda }F_{\mu \nu }F^{\mu
\nu }$. To get the term, $\xi _{\mu }\xi _{\nu }F{^{\mu }}_{\kappa
}F^{\kappa \nu }$, we could also propose the action below:

\begin{equation}
S=\kappa \int d^{4}xd^{2}\theta d^{2}\bar{\theta}\left\{ \left( D^{\alpha
}\Xi \right) (D_{\alpha }\Xi )W^{\beta }W_{\beta }+h.c.\right\} .  \label{30}
\end{equation}

Also in this case, when we supersymmetrise the $\xi _{\mu }\xi _{\nu }F{%
^{\mu }}_{\kappa }F^{\kappa \nu }$-term, automatically comes out $\xi
_{\lambda }\xi ^{\lambda }F^{2}$. So, we have a second way to build up the
supersymmetric extension of the term $\left( \frac{1}{2}\xi _{\mu }\xi _{\nu
}F{^{\mu }}_{\kappa }F^{\kappa \nu }+\frac{1}{8}\xi _{\lambda }\xi ^{\lambda
}F_{\mu \nu }F^{\mu \nu }\right) $. We just mention this possible second way
of working out the supersymmetrisation of (\ref{acao2}) by a vector
superfield background, but we do not exploit here this second possibility.
This is why we do not project \ the superfield action (\ref{30}) into
components.

Though we formulate our model in terms of superspace and superfields, the
tensor calculus of supersymmetry, we would like to point out \ that with a
Lorentz-symmetry breaking, in the chiral case given by $\partial _{\mu
}S\neq 0$, in the vector case parametrized by $\xi ^{\lambda }\neq 0$, SUSY
is readily seen to be broken by the background. But, as it was emphasized,
we do not have elements to decide whether SUSY breaking has occurred before
Lorentz-symmetry violation, though the breaking of the latter signals the
violation of the former.

\section{Concluding Comments}

Our effort in this paper has mainly consisted in finding out possible $N=1$
supersymmetric scenarios for the $K-$ tensor- realized Lorentz-symmetry
breaking considered in the literature ( see the Refs. quoted throughout the
present work).

We propose two viable descriptions. The first approach is based on a chiral
superfield that accommodates the content of the background responsible for
the breaking of Lorentz-symmetry. In this case, a complex scalar is the
source for $K_{\mu \nu \kappa \lambda }$. The constancy of the $K-$ tensor
is ensured by the linear dependence of $S$ on the $x^{\mu }-$ coordinates.
An interesting question to be investigated is the analysis of the photon -
photino mass splitting in terms of the background field components,
specially if the fermion component $\zeta $ is a non-vanishing constant
background. From the expression for $S_{coupled}$ in eq. (\ref{smix1}),
there result interesting terms that mix the gauge superfield components ($%
A^{\mu },\lambda ,$ and $D$) \ in bilinear terms where the (constant)
background fields are also present, as mass parameters. The task of getting
the $<A_{\mu }A_{\nu }>-$ $<A_{\mu }\lambda >-<\lambda \lambda >-<\lambda
D>- $ and $<DD>-$propagators demands special technicalities ( Fierzings and
spin--projection operators) and, once these tree level $2-$ point functions
are worked out, we can read off the propagator poles and discuss in terms of
the background components ( $S,$ $\zeta ,$ and $G$). The r\^{o}le of the
background fermionic condensates become clear after the propagator poles are
identified.

This discussion holds through also in the cases of a vector-superfield
background that carries $\xi _{\mu }$, from which $K_{\mu \nu \kappa \lambda
}$ is expressed. This case exhibits a richer background and fermion
condensates that mix $\chi $ and $\Psi $ ( see eq. (\ref{smix2})), in
addition to the $\chi -$ and $\Psi -$condensates become important.

So, to our mind, in either case, the important question that our study may
raise concerns the influence of the Lorentz-symmetry violating background on
the spectrum and which restrictions it should have so as to avoid the
appearance of non-physical excitations, such as tachyons and ghosts.

Finally, a non-trivial question that remains to be addressed to is the
relation between SUSY and Lorentz-symmetry breakings. We treat the latter in
a supersymmetric formulation and, as we have previously commented, our
backgrounds (both $\Omega $ \ and $\Xi $)\ are not invariant under SUSY.
They simply express the fact that Lorentz-symmetry breaking does not support
exact SUSY. However, by no means, we are stating that we break
Lorentz-symmetry and SUSY at the same time. SUSY could have been broken
before, by some more fundamental mechanism, and Lorentz-symmetry breaking
takes place in an environment which still keeps the inheritance of SUSY
through the whole set of supersymmetric partners. This is why approach the
deviation from Lorentz-symmetry in connection with SUSY. However, it would
be an interesting task- and we shall soon concentrate on this point - to
build up a model such that, whenever SUSY is spontaneously broken, some
tensor field (that belongs to some supermultiplet of the model coupled to
the superfield responsible for the breaking of SUSY) also acquires a
non-trivial vacuum expectation value and the Lorentz \ group simultaneously
undergoes spontaneous symmetry breakdown. This sets out another possibility,
with a contemporary breaking of SUSY and Lorentz-symmetry. We are
considering this situation and we shall report on it elsewhere.

\section{Acknowledgments}

The authors F. J. L. Leal and J. A. Helay\"{e}l-Neto thank the Brazilian
funding agencies, CNPq and FAPERJ for financial support.

\end{document}